\documentclass[plb,onecolumn,nofootinbib,showpacs,floatfix,12pt]{revtex4}
\usepackage{graphicx}
\usepackage{amsmath,amssymb,bm,dcolumn}
\usepackage[export]{adjustbox}
\usepackage{graphicx}
\usepackage{float}
\usepackage{amsmath}
\usepackage{amsfonts}
\usepackage{color}
\usepackage{slashed}
\usepackage[colorlinks,hyperindex]{hyperref}
\usepackage{epsfig}		
\usepackage{dcolumn}
\usepackage{bm}
\usepackage{multirow}
\usepackage{longtable}
\usepackage{array}
\usepackage{booktabs}
\usepackage{array} 
\usepackage{varwidth} 

\hypersetup{
colorlinks,
citecolor=blue,
linkcolor=red,
urlcolor=black,
}
\def\be{\begin{equation}}
\def\ee{\end{equation}}
\def\bed{\begin{description}}
\def\eed{\end{description}}
\def\bea{\begin{eqnarray}}
\def\eea{\end{eqnarray}}

\def\ba{\begin{array}}
\def\ea{\end{array}}

\def\u1{$U(1)$}
\def\suu1{$SU(2)\times U(1)$}

\def\0{\mbox{\tiny $0$}}
\def\1{\mbox{\tiny $1$}}
\def\2{\mbox{\tiny $2$}}
\def\3{\mbox{\tiny $3$}}
\def\4{\mbox{\tiny $4$}}
\def\5{\mbox{\tiny $5$}}
\def\6{\mbox{\tiny $6$}}
\def\7{\mbox{\tiny $7$}}
\def\8{\mbox{\tiny $8$}}
\def\9{\mbox{\tiny $9$}}

\def\f14{\mbox{\tiny $\frac{1}{4}$}}

\DeclareMathOperator{\sech}{sech}

\newcolumntype{L}[1]{>{\raggedright\let\newline\\\arraybackslash\hspace{0pt}}m{#1}}
\newcolumntype{C}[1]{>{\centering\let\newline\\\arraybackslash\hspace{0pt}}m{#1}}
\newcolumntype{R}[1]{>{\raggedleft\let\newline\\\arraybackslash\hspace{0pt}}m{#1}}

\begin{document}

\title{Supersymmetry and Fermionic Modes in an Oscillon Background}
\author{R. A. C. Correa$^{1,}$}
\email{fis04132@gmail.com}
\author{L. P. R. Ospedal$^{2,}$}
\email{leoopr@cbpf.br}
\author{W. de Paula$^{1,}$}
\email{wayne@ita.br}
\author{J. A. Helay\"el-Neto$^{2,}$}
\email{helayel@cbpf.br}
\affiliation{\footnotesize{$^1$ITA, Instituto Tecnol\'ogico de Aeron\'autica, 12228-900, S\~ao Jos\'e
dos Campos, SP, Brazil}}
\affiliation{\footnotesize{$^2$CBPF, Centro Brasileiro de Pesquisas F\'{i}sicas, 22290-180, Rio de Janeiro, RJ, Brazil}}
\date{\today}

\begin{abstract}{The excitations referred to as oscillons are long-lived time-dependent field configurations which emerge dynamically from non-linear field theories. Such long-lived solutions are of interest in applications that include systems of Condensed Matter Physics, the Standard Model of Particle Physics, Lorentz-symmetry violating scenarios and Cosmology.  In this work, we show how oscillons may be accommodated  in a supersymmetric scenario. We adopt as our framework simple ($\mathcal{N}=1$) supersymmetry  in $D=1+1$ dimensions. We focus on the bosonic sector with oscillon configurations and their (classical) effects on the corresponding fermionic modes, (supersymmetric) partners of the oscillons. The particular model we adopt to pursue our investigation displays cubic {\bf superfield which, in the physical scalar sector, corresponds to the usual quartic self-coupling.} }

\end{abstract}

\date{\today}
\maketitle

\section{Introduction}

\label{sec:intro} The study of field-theoretic non-linear systems is an area of
increasing interest over the past few decades \cite{Whitham, Scott, Weinberg}. Evidences of non-linear behavior is, nowadays, found in a considerable 
part of physical systems. This includes systems of Condensed Matter
Physics, field-theoretic models related to particle phenomenology, modern Cosmology 
and a large number of other domains of the physical sciences \cite{Rajaraman}-\cite{Segur1}. \textcolor{black}{As examples, we can cite studies on nonlinear acoustic metamaterials (NAMs) \cite{FangNature2017}, in the resonances of two-dimensional (2D) materials \cite{DavidovikjNature2017}, in nonlinear sigma models \cite{DanielJHEP2017}, in hard x-ray pulses in free-electron lasers (FELs) \cite{DingPRL2017}, and in supersymmetric quantum field theories with kink excitations \cite{MussardoJHEP2007}.}
 

In classical field theory, there is a class of configuration quite common and important called
solitons \cite{vachaspati}, which are solutions of a set of classical relativistic non-linear field equations. Such configurations are characterized by some topological index, related to their behavior at spatial infinity. Solitons have the important feature of having energy density localized at space and having its profiles restored to their original shapes and velocities after collisions. Nowadays, those configurations is well understood in a broad number of scenarios. For instance, one can find investigations regarding monopoles, textures,
strings and kinks \cite{Vilenkin}. 

Within the universe of non-linear field theories, it is important to highlight a specially important class of time-dependent
stable solutions referred to in the literature as breathers. This configurations comes from Sine-Gordon like models.
Another time-dependent field configuration whose stability is granted for by
charge conservation are the $Q$-balls \ as named by Coleman \cite%
{S.Coleman1} or non-topological solitons, according to Lee \cite{Lee}. On the other hand, besides these physical systems that exhibit a
metastable behavior, a new class of solutions was discovered in the years 1970 by Bogolyubsky
and Makhankov \cite{I.L.Bogolyubsky}, and re-assessed afterwards by
Gleiser \cite{Gleiser1}. These solutions were identified during the study of the
dynamics of first-order phase transitions, where oscillons arise from collapsing
unstable spherically-symmetric bubbles in models with symmetric and asymmetric double-well potentials. Since that work, a series of investigations is addressed 
to study these objects \cite{Gleiser2}-\cite{Hertzberg}. {For instance, we can find interesting investigations and consequences in Abelian-Higgs models \cite{adc1,adc2}, in massive Yang-Mills theories \cite{adc3}, and when there are nonlinear Schrodinger equations \cite{adc4}.}

Oscillons do not keep their shape indefinitely. Indeed, for a $\phi^4$ theory in $(1+1)$ dimension, Segur and Kruskal \cite{Segur1} have shown that oscillon 
configurations slowly emit their energy. A similar conclusion was obtained in the cases of $(2+1)$ and $(3+1)$ dimensions \cite{Gyula2}. 
Also, the radiation rate for quantized oscillons was worked out by Hertzberg \cite{Hertzberg}, who showed that there is
a significant difference between the classical and quantum decays.

The most common oscillon profile is a bell-like shape with a sinusoidal oscillation. A quite different oscillon configuration 
may be found in Ref. \cite{mustafa}. In that work, one analyzes the properties of oscillons in an expanding Universe. Interestingly, a new kind of oscillon
which presents a plateau at its top is found, and cosmological applications can be implemented. It is also shown that these configurations present a stable behavior against 
localized perturbations.

 In this contribution, we endeavour to present a (1+1)-dimensional supersymmetric framework
where oscillon configurations show up and it is our interest to focus on the oscillon fermionic partner,
that we may refer to as oscillino. The reason to consider oscillons in connection with supersymmetry
(from now on, SUSY) is not of a purely academic interest. We take the viewpoint that fermions are  the truly elementary 
matter excitations of space-time. Associating SUSY with oscillons is a natural way to couple oscillons to fermions if we
wish to inspect their interaction. In so doing, besides the interest in identifying the profile of the fermionic configurations coupled to 
oscillons (which we do by building up oscillino-type solutions), it would be interesting to consider the behavior of oscillino condensates 
to find out whether it is possible to identify them with scalar or vector bosonic excitations present in lower-dimensional systems. 
Also, in a SUSY scenario, oscillino condensates, in turn, induce correction effects on the oscillons, and it might be interesting to assess 
how these effects may interfere on the stability of the oscillons. These are the primary motivations underneath our proposal to inspect oscillons 
in connection with SUSY.

This work is organized as follows: in Section $2$, we introduce the general features of the supersymmetric model. After that, in Section $3$, we review the 
oscillon configurations in the purely bosonic sector, with SUSY not yet switched on. Next, in Section $4$, we  investigate the fermionic partner's solutions in 
an oscillon background and we discuss on two different paths to get oscillino solutions. Finally, we present our Concluding Comments in Section $5$. An Appendix
follows, where we set up conventions and present the off-shell version of the SUSY action we adopt to carry out our inspections.


\section{The supersymmetric model}

\label{sec:mod}

 The investigation of supersymmetric models in the context of nontrivial classical field configurations finds a remarkable result in the paper by Witten and Olive \cite{Witten_Olive}, 
where a connection between topological configurations and central charges of the SUSY algebra is established. Furthemore, many other field-theoretic models have been discussed
in this context. Specifically, in $D=1+1$ dimensions, we highlight a number of seminal works \cite{Vecchia_Ferrara}-\cite{Adda}, which explored non-pertubative classical solutions and non-linear sigma-models. We also point out some works \cite{Naculich}-\cite{Aguirre}, where the quantum aspects of (non-)topological solutions in a supersymmetric scenario are discussed. However, one notices a lack of the attention related to oscillon configurations and supersymmetry. Here, we present a first contribution to this topic.  In this Section, the Lagrangian density describing a theory in $D=1+1$
dimensions in a supersymmetric framework is introduced. Our aim in working
with a supersymmetric theory comes from the fact that, in its linear
realization, SUSY provides the existence of a new class of field
configurations, which correspond to the supersymmetric partner of the
oscillons, named from now on \textit{oscillino}. The on-shell $\mathcal{N}=1$
supersymmetric Lagrangian density is given by

\begin{equation}
\mathcal{L} = \frac{1}{2}(\partial _{\mu }\varphi )^{2}+\frac{i}{2}\bar{\psi}%
\gamma ^{\mu }\partial _{\mu }\psi -\frac{[V_{\varphi }(\varphi )]^{2}}{2}-%
\frac{\bar{\psi}\psi }{2}V_{\varphi \varphi }(\varphi ),  \label{e1}
\end{equation}

\noindent where $\varphi $ and $\psi $ are, respectively, a real scalar field 
and a Majorana spinor. Here $V(\varphi )$, an arbitrary function, {\bf is referred to as the prepotential,} and we
denote $V_{\varphi }\equiv \partial V/\partial \varphi $ and $V_{\varphi
\varphi }\equiv \partial ^{2}V/\partial \varphi ^{2}$. Moreover, we
highlight that the scalar potential is related to $V(\varphi )$ {\bf as below:}

\begin{equation}
U(\varphi )=\frac{1}{2}[V_{\varphi }(\varphi )]^{2}.
\label{potential} \end{equation}

In the Appendix A, we present our conventions as well as the
off-shell Lagrangian density obtained in the superspace formulation.

One can check that the action is invariant under the transformations

\begin{eqnarray}
\delta \varphi &=& \bar{\xi} \psi \, ,  \label{on_transf_varphi} \\
\delta \psi &=& -i \, \gamma^\mu \xi \, \partial_\mu \varphi - \xi \,
V_{\varphi} \, ,  \label{on_transf_psi}
\end{eqnarray}

\noindent where $\xi $ is a Majorana spinor parameter.

From the Lagrangian density (\ref{e1}), we obtain the following coupled
equations of motion\textbf{\ }

\begin{eqnarray}
\square \varphi +V_{\varphi }V_{\varphi \varphi }+\frac{\bar{\psi}\psi }{2}%
V_{\varphi \varphi \varphi } &=&0,  \label{e4} \\
i\gamma ^{\mu }\partial _{\mu }\psi -\psi V_{\varphi \varphi } &=&0.
\label{e5}
\end{eqnarray}

In this work, we investigate the particular case

\begin{equation}
V(\varphi )=\frac{a}{3}\varphi ^{3}-\frac{b}{2}\varphi ^{2}.  \label{e6}
\end{equation}

\noindent where $a$ and $b$ are real (positive) parameters. {\bf $V(\varphi)$ is cubic, so that the scalar self-interaction becomes quartic, as given in Eq. \eqref{potential}.}

In the sequel, we turn our attention to the problem of how to proceed to
decouple the equations. In order to solve the equations to obtain oscillon-type
configurations, we consider, as our initial step, the approximation that the  interaction with the fermionic condensate $(\bar{\psi} \psi)$ can be neglected. 
 One possible situation where this consideration is not an approximation is the case of half-SUSY, where the spinor field is Majorana-Weyl; but, this is not the case here. 
We start with the scalar field behaving like a classical background. A non-trivial fermionic solution shall be obtained by solving Eq. (\ref{e5}) in the oscillon background or, equivalently, 
by perturbing the oscillon configuration by means of a SUSY transformation, (Eqs. (\ref{on_transf_varphi}) and (\ref{on_transf_psi})). Thus, Eq. (\ref{e4})
becomes

\begin{equation}
\square \varphi +V_{\varphi }V_{\varphi \varphi }=0.  \label{e7}
\end{equation}

Using the Eq. (\ref{e6}), we can rewrite the Eq. (\ref{e7}) as follows below:

\begin{equation}
\square \varphi +2a^{2}\varphi ^{3}-3ab\varphi ^{2}+b^{2}\varphi =0.
\label{e8}
\end{equation}

As one notices, the equation above involves only the scalar sector. In the
next Section, we shall work out the oscillons configurations from this
scenario.


\section{Bosonic sector: The usual oscillons}

\label{sec:bsec}

Since our first interest is to find periodic and localized solutions, it is
useful to introduce a scale
transformations in $x$ and $t$,  given by

\begin{equation}
\tau =\omega t,\text{ \ }y=\epsilon x,  \label{e9}
\end{equation}

\noindent with $\omega=\sqrt{1-\epsilon^2}$ and $0<\epsilon\leq1$. Thus, Eq. (\ref{e8}) becomes

\begin{equation}
\omega ^{2}\partial _{\tau }^{2}\varphi -\epsilon ^{2}\partial
_{y}^{2}\varphi +2a^{2}\varphi ^{3}-3ab\varphi ^{2}+b^{2}\varphi =0.
\label{e10}
\end{equation}

From  this equation, it is possible to find a oscillon configuration, which
are localized in the central vacuum $\varphi _{v}=0$ of the model described
in Eq. (\ref{e6}). In this case, the classical scalar field $\varphi $ is
spatially localized and periodic in time. The usual procedure to obtain
oscillon configurations in $D=1+1$ dimensions consists in applying a small
amplitude expansion of the scalar field in powers of $%
\epsilon$ in the form that follows: 

\begin{equation}
\varphi (y,\tau )=\sum\limits_{j=1}^{\infty }\epsilon ^{j}\varphi
_{j}(y,\tau ).  \label{e11}
\end{equation}

Let us replace this expansion of the scalar field into the field equation
(\ref{e10}). Doing that yields:

\begin{eqnarray}
&&\left. \epsilon \left( \partial _{\tau }^{2}\varphi _{1}+b^{2}\varphi
_{1}\right) +\epsilon ^{2}\left( \partial _{\tau }^{2}\varphi
_{2}+b^{2}\varphi _{2}-3ab\varphi _{1}\right) \right.  \notag \\
&&\left. +\epsilon ^{3}\left( \partial _{\tau }^{2}\varphi _{3}+b^{2}\varphi
_{3}-\partial _{\tau }^{2}\varphi _{1}-\partial _{y}^{2}\varphi
_{1}-6ab\varphi _{1}\varphi _{2}+2a^{2}\varphi _{1}^{3}\right) +...=0.\right.
\label{e12}
\end{eqnarray}

We notice that the procedure of performing a small amplitude expansion shows
that the scalar field solution $\varphi$ can be obtained from a set of
scalar fields which satisfy coupled non-linear differential equations. This
set of differential equations is found by taking the terms to all orders in $%
\epsilon$ in the above equation. Thus, it becomes immediately clear that, up
to $\epsilon^3$, this procedure leads to

\begin{eqnarray}
\partial _{\tau }^{2}\varphi _{1}+b^{2}\varphi _{1} &=&0,  \label{e13} \\
\partial _{\tau }^{2}\varphi _{2}+b^{2}\varphi _{2}-3ab\varphi _{1}^2 &=&0,
\label{e14} \\
\partial _{\tau }^{2}\varphi _{3}+b^{2}\varphi _{3}-\partial _{\tau
}^{2}\varphi _{1}-\partial _{y}^{2}\varphi _{1}-6ab\varphi _{1}\varphi
_{2}+2a^{2}\varphi _{1}^{3} &=&0.  \label{e15}
\end{eqnarray}

 From Eq. (\ref{e13}), we can propose that

\begin{equation}
\varphi _{1}(y,\tau )=\Phi \left( y\right) \cos \left( b\tau \right) ,
\label{e16}
\end{equation}

\noindent where $\Phi \left( y\right) $ is function of the spatial variable $%
y$. We notice that the lowest order term of the solution is just a harmonic
oscillator in time, with frequency $b$.

On the other hand, the solution to the Eq. (\ref{e14}), which is a linear
inhomogeneous equation, can be found by considering the Eq. (\ref{e16}).
Thus, the solution for $\varphi _{2}(y,\tau )$ is written as below:

\begin{equation}
\varphi _{2}(y,\tau )=\frac{a\Phi \left( y\right) ^{2}}{2 b}\left[ 3-\cos
\left( 2b\tau \right) \right] .  \label{e17}
\end{equation}

From these solutions, Eqs. (\ref{e16}) and (\ref{e17}), we can obtain $%
\varphi _{3}(y,\tau )$. Then, after straightforward calculations, one can
verify that Eq. (\ref{e15}) takes the form

\begin{equation}
\partial _{\tau }^{2}\varphi _{3}+b^{2}\varphi _{3}=\left( \Phi "-b^{2}\Phi
+6a^{2}\Phi ^{3}\right) \cos \left( b\tau \right) -2a^{2}\Phi ^{3}\cos
\left( 3b\tau \right) .  \label{e18}
\end{equation}
where $\Phi " = d^2 \Phi/dy^2$. 

Our aim is to get configurations which are periodical in time. Then, if we
solve the above partial differential equation in the presented form, we will
have a term linear in $\tau $. As a consequence, the solution for $\varphi
_{3}$ is neither periodical nor localized. This result comes from the
contribution of the function $\cos \left( b\tau \right) $ in the right-hand
side of the partial differential equation (\ref{e18}). However, we can
build up solutions for $\varphi _{3}$ which are periodical in time if we
impose that

\begin{equation}
\Phi "-b^{2}\Phi +6a^{2}\Phi ^{3}=0.  \label{e19}
\end{equation}

In this case one gets

\begin{equation}
\Phi \left( y \right) =\frac{b}{a\sqrt{3}} \sech \left( by \right) .  \label{e20}
\end{equation}

From the above results, as one can see, the leading order corresponding solution for the classical field is given by

\begin{equation}
\varphi (x,t)\simeq \frac{\epsilon b}{a\sqrt{3}}\sech\left( \epsilon
bx\right) \cos \left( b\sqrt{1-\epsilon ^{2}}t\right)  .  \label{e21}
\end{equation}

We note that the bosonic sector accommodates usual oscillons configurations
with small amplitude.

{ \bf Finally, it is worthy to emphasize that we have expanded the scalar field up to 3th order in $\epsilon$, obtaining thereby a system of equations (Eqs. \eqref{e13}-\eqref{e15}). The solutions to this system have a localized spatial profile with a $\sech (by)$-functional dependence; the leading contribution is therefore the term proportional to $\epsilon$. Notice that it is necessary to use the expansion up to 3th order in $\epsilon$ to obtain Eq. \eqref{e19}, whose solution determines the function $\Phi(y)$, which is responsible for the spatial profile of the oscillons. We then present the leading-order contribution for the scalar, given by Eq. \eqref{e21}.  Also, in Ref. \cite{Hertzberg}, it is shown that the leading contribution for the oscillon radiation is of order $\epsilon$, so that the other terms are subleading for the analysis of the oscillon stability. A similar procedure shall be adopted in the fermionic case, in dealing with the oscillino solution. }

In the next Section, we shall study the fermionic sector in the background of this oscillon.


\section{The SUSY partners of oscillons}

\label{sec:fsec}

 Having established the oscillon configuration, we now turn our attention to the fermionic sector. Our goal is to find the
supersymmetric partner of the  oscillon, which we shall refer to as \textit{oscillino}. To do this, let us start by
rewriting the Dirac equation (\ref{e5})  in the presence of the potential  (\ref{e6}). In this case, we have
\begin{equation}
i\gamma ^{\mu }\partial _{\mu }\psi -(2a\varphi -b)\psi =0.  \label{eq1.1}
\end{equation}

As we are working with two dimensions, the fermions are described by
two-component spinors, 

\begin{equation}
\psi (x,t)=\left( 
\begin{tabular}{c}
$\psi _{1}(x,t)$ \\ 
$\psi _{2}(x,t)$%
\end{tabular}%
\right) .  \label{eq1.2}
\end{equation}

 As already mentioned in the Appendix, we adopted the Majorana  representation of the gamma matrices $(\gamma ^{0}=\sigma _{y}\text{, }\gamma ^{1}=i\sigma _{x})$.  Therefore, by using this representation and Eqs. (\ref{eq1.1}) and (\ref{eq1.2}) one  can arrive at following coupled pair of first order differential equations:

\begin{eqnarray}
\partial _{t}\psi _{2}+\partial _{x}\psi _{2}-S(\varphi )\psi _{1} &=&0,
\label{eq1.3} \\
-\partial _{t}\psi _{1}+\partial _{x}\psi _{1}-S(\varphi )\psi _{2} &=&0,
\label{eq1.4}
\end{eqnarray}

\noindent  where $S(\varphi )\equiv 2a(\varphi -b/2a)$.

Now, with the purpose of decouple these equations,
it is necessary to substitute the Eq.~(\ref{eq1.3}) into (\ref{eq1.4}) and 
vice versa,  which leads to the
corresponding second order differential equations:

\begin{eqnarray}
\square \psi _{1}+\frac{S_{\varphi }(\varphi )}{S(\varphi )}\left[ (\partial
_{t}\varphi +\partial _{x}\varphi )(-\partial _{t}\psi _{1}+\partial
_{x}\psi _{1})\right] +S^{2}(\varphi )\psi _{1} &=&0,  \label{eq1.7} \\
\square \psi _{2}-\frac{S_{\varphi }(\varphi )}{S(\varphi )}\left[ (\partial
_{t}\varphi -\partial _{x}\varphi )(\partial _{t}\psi _{2}+\partial _{x}\psi
_{2})\right] +S^{2}(\varphi )\psi _{2} &=&0.  \label{eq1.8}
\end{eqnarray}

As we are interested in obtaining the supersymmetric oscillons, it is
natural to apply the scales transformation in $x$ and $t$ introduced in Eq. (%
\ref{e9}). Thus, the above equations can be rewritten in the form%
\begin{eqnarray}
\omega ^{2}\partial _{\tau }^{2}\psi _{1}-\epsilon ^{2}\partial _{y}^{2}\psi
_{1}+\frac{S_{\varphi }}{S}\left[ (\omega \partial _{\tau }\varphi + \epsilon
\partial _{y}\varphi )(-\omega \partial _{\tau }\psi _{1}+\epsilon \partial
_{y}\psi _{1})\right] +S^{2}(\varphi )\psi _{1} &=&0,  \label{eq1.9} \\
\omega ^{2}\partial _{\tau }^{2}\psi _{2}-\epsilon ^{2}\partial _{y}^{2}\psi
_{2}-\frac{S_{\varphi }}{S}\left[ (\omega \partial _{\tau }\varphi -\epsilon
\partial _{y}\varphi )(\omega \partial _{\tau }\psi _{2}+\epsilon \partial
_{y}\psi _{2})\right] +S^{2}(\varphi )\psi _{2} &=&0.  \label{eq2.0}
\end{eqnarray}

After that, let us use the small amplitude expansion for the fields $\psi _{1}$ and 
$\psi _{2}$. Here, we will assume that%
\begin{eqnarray}
\psi _{1}(y,\tau ) &=&\sum\limits_{j=1}^{\infty }\epsilon ^{j}\sigma
_{j}(y,\tau ),  \label{eq2.1} \\
\psi _{2}(y,\tau ) &=&\sum\limits_{j=1}^{\infty }\epsilon ^{j}\rho
_{j}(y,\tau ).  \label{eq2.2}
\end{eqnarray}

By replacing the expansion given above into Eqs. (\ref{eq1.9}) and (\ref{eq2.0}),
we obtain%
\begin{eqnarray}
&&\left. \epsilon (\partial _{\tau }^{2}\sigma _{1}+b^{2}\sigma
_{1})+\epsilon ^{2}\left[ \partial _{\tau }^{2}\sigma _{2}+b^{2}\sigma _{2}-%
\frac{4b^{2}\cos (b\tau )\sech(by)\sigma _{1}}{\sqrt{3}}-\frac{2b\sech%
(by)\sin (b\tau )\partial _{\tau }\sigma _{1}}{\sqrt{3}}\right] \right. 
\notag \\
&&\left. +\epsilon ^{3}\left[ \partial _{\tau }^{2}\sigma _{3}+b^{2}\sigma
_{3}+\frac{4}{3}b^{2}\cos ^{2}\left( b\tau \right) \sech^{2}(by)\sigma _{1}-%
\frac{4b^{2}\cos (b\tau )\sech(by)\sigma _{2}}{\sqrt{3}}\right. \right. 
\notag \\
&&\left. -\frac{4}{3}b\cos \left( b\tau \right) \sech^{2}(by)\sin (b\tau
)\partial _{\tau }\sigma _{1}+\frac{2b\cos (b\tau )\sech(by)\tanh
(by)\partial _{\tau }\sigma _{1}}{\sqrt{3}}\right.  \notag \\
&&\left. \left. -\frac{2b\sech(by)\sin (b\tau )\partial _{\tau }\sigma _{2}}{%
\sqrt{3}}-\partial _{\tau }^{2}\sigma _{1}+\frac{2b\sech(by)\sin (b\tau
)\partial _{y}\sigma _{1}}{\sqrt{3}}-\partial _{y}^{2}\sigma _{1}\right]
+O(\epsilon ^{4})=0,\right.  \label{eq2.3} \\
&&  \notag \\
&&\left. \epsilon (\partial _{\tau }^{2}\rho _{1}+b_{1}^{2}\rho )+\epsilon
^{2}\left[ \partial _{\tau }^{2}\rho _{2}+b^{2}\rho _{2}-\frac{4b^{2}\cos
(b\tau )\sech(by)\rho _{1}}{\sqrt{3}}-\frac{2b\sech(by)\sin (b\tau )\partial
_{\tau }\rho _{1}}{\sqrt{3}}\right] \right.  \notag \\
&&\left. +\epsilon ^{3}\left[ \partial _{\tau }^{2}\rho _{3}+b_{3}^{2}\rho
_{3}+\frac{4}{3}b^{2}\cos ^{2}\left( b\tau \right) \sech^{2}(by)\rho _{1}-%
\frac{4b^{2}\cos (b\tau )\sech(by)\rho _{2}}{\sqrt{3}}\right. \right.  \notag
\\
&&\left. -\frac{4}{3}b\cos \left( b\tau \right) \sech^{2}(by)\sin (b\tau
)\partial _{\tau }\rho _{1}+\frac{2b\cos (b\tau )\sech(by)\tanh (by)\partial
_{\tau }\rho _{1}}{\sqrt{3}}\right.  \notag \\
&&\left. \left. -\frac{2b\sech(by)\sin (b\tau )\partial _{\tau }\rho _{2}}{%
\sqrt{3}}-\partial _{\tau }^{2}\rho _{1}-\frac{2b\sech(by)\sin (b\tau
)\partial _{y}\rho _{1}}{\sqrt{3}}-\partial _{y}^{2}\rho _{1}\right]
+O(\epsilon ^{4})=0.\right.  \label{eq2.6}
\end{eqnarray}

The procedure of performing a small amplitude expansion shows that the
fields $\psi _{1}$- and $\psi _{2}$-fields can be obtained from a
set of  fields which satisfy coupled non-linear differential equations.
This set of differential equations is found by taking the terms in all
orders of $\epsilon $ in the above equation. Thus,  one can check that up to $\epsilon ^{3}$ the above supposition leads to the
following set of equations

\begin{eqnarray}
&&\left. \partial _{\tau }^{2}\sigma _{1}+b^{2}\sigma _{1}=0,\right.
\label{1} \\
&&\left. \partial _{\tau }^{2}\rho _{1}+b^{2}\rho _{1}=0,\right.  \label{2}
\end{eqnarray}

\begin{eqnarray}
&&\left. \partial _{\tau }^{2}\sigma _{2}+b^{2}\sigma _{2}-\frac{4b^{2}\cos
(b\tau )\sech(by)\sigma _{1}}{\sqrt{3}}-\frac{2b\sech(by)\sin (b\tau
)\partial _{\tau }\sigma _{1}}{\sqrt{3}}=0,\right.  \label{3} \\
&&\left. \partial _{\tau }^{2}\rho _{2}+b^{2}\rho _{2}-\frac{4b^{2}\cos
(b\tau )\sech(by)\rho _{1}}{\sqrt{3}}-\frac{2b\sech(by)\sin (b\tau )\partial
_{\tau }\rho _{1}}{\sqrt{3}}=0,\right.  \label{4}
\end{eqnarray}

\begin{eqnarray}
&&\left. \partial _{\tau }^{2}\sigma _{3}+b^{2}\sigma _{3}+\frac{4}{3}%
b^{2}\cos ^{2}\left( b\tau \right) \sech^{2}(by)\sigma _{1}-\frac{4b^{2}\cos
(b\tau )\sech(by)\sigma _{2}}{\sqrt{3}}\right.  \notag \\
&&\left. -\frac{4}{3}b\cos \left( b\tau \right) \sech^{2}(by)\sin (b\tau
)\partial _{\tau }\sigma _{1}+\frac{2b\cos (b\tau )\sech(by)\tanh
(by)\partial _{\tau }\sigma _{1}}{\sqrt{3}}\right.  \notag \\
&&\left. -\frac{2b\sech(by)\sin (b\tau )\partial _{\tau }\sigma _{2}}{\sqrt{3%
}}-\partial _{\tau }^{2}\sigma _{1}+\frac{2b\sech(by)\sin (b\tau )\partial
_{y}\sigma _{1}}{\sqrt{3}}-\partial _{y}^{2}\sigma _{1}=0,\right.  \label{5}
\end{eqnarray}

\begin{eqnarray}
&&\left. \partial _{\tau }^{2}\rho _{3}+b_{3}^{2}\rho _{3}+\frac{4}{3}%
b^{2}\cos ^{2}\left( b\tau \right) \sech^{2}(by)\rho _{1}-\frac{4b^{2}\cos
(b\tau )\sech(by)\rho _{2}}{\sqrt{3}}\right.  \notag \\
&&\left. -\frac{4}{3}b\cos \left( b\tau \right) \sech^{2}(by)\sin (b\tau
)\partial _{\tau }\rho _{1}+\frac{2b\cos (b\tau )\sech(by)\tanh (by)\partial
_{\tau }\rho _{1}}{\sqrt{3}}\right.  \notag \\
&&\left. -\frac{2b\sech(by)\sin (b\tau )\partial _{\tau }\rho _{2}}{\sqrt{3}}%
-\partial _{\tau }^{2}\rho _{1}-\frac{2b\sech(by)\sin (b\tau )\partial
_{y}\rho _{1}}{\sqrt{3}}-\partial _{y}^{2}\rho _{1}=0.\right.  \label{eq3.3}
\end{eqnarray}

From now on, we shall solve these equations. {\bf  For the sake of simplicity, let us propose a particular solution for the first two equations, namely,}

\begin{eqnarray}
\sigma _{1}(y,\tau ) &=&\sigma (y)\cos (b\tau ),  \label{eq3.4} \\
\rho _{1}(y,\tau ) &=&\rho (y)\sin (b\tau ).  \label{eq3.5}
\end{eqnarray}

Plugging the solutions (\ref{eq3.4}) and (\ref{eq3.5}) into Eqs. (\ref{3}%
) and (\ref{4}), we have

\begin{eqnarray}
\sigma _{2}(y,\tau ) &=&-\frac{2\sigma (y)\left[ -3+\cos (2b\tau )\right] %
\sech(by)}{3\sqrt{3}},  \label{eq3.6} \\
\rho _{2}(y,\tau )&=&-\frac{\rho (y)}{\sqrt{3}}\sech\left( by\right)
\sin (2b\tau ).  \label{eq3.7}
\end{eqnarray}

Using the results (\ref{eq3.4})-(\ref{eq3.7}), the Eqs. (\ref{5}) and (\ref%
{eq3.3}) can be rewritten as

\begin{eqnarray}
&&\left. \partial _{\tau }^{2}\sigma _{3}+b^{2}\sigma _{3}=-\frac{%
8b^{2}\sigma (y)\cos \left( b\tau \right) \left[ -3+\cos (2b\tau )\right] %
\sech^{2}(bx)}{9}\right.   \notag \\
&&\left. -\frac{b\sech(by)\sin (2b\tau )\sigma ^{\prime }(y)}{\sqrt{3}}+\cos
(b\tau )\sigma ^{\prime \prime }(y)\sigma (y)\right.  \\
&&-\frac{1}{18}b^{2}\cos (b\tau )\sech^{2}(by)\left[ 17+16\cos (2b\tau
)+9\cosh (2by)-12\sqrt{3}\sin (b\tau )\sinh (by)\right] \sigma (y),  \notag
\end{eqnarray}

\begin{eqnarray}
\partial _{\tau }^{2}\rho _{3}+b^{2}\rho _{3} &=&-b^{2}\rho (y)\sin (b\tau
)-b^{2}\rho (y)\cos (2b\tau )\sech^{2}(by)\sin (b\tau ) \notag \\
&&-\frac{4}{3}b^{2}\rho (y)\cos (b\tau )\sech^{2}(by)\sin (2b\tau ) \notag \\
&&-\frac{2}{\sqrt{3}}b^{2}\rho (y)\cos ^{2}(b\tau )\sech(by)\tanh
(by)\sin (b\tau )+\rho^{\prime \prime }(y)\sin (b\tau ) \\
&&+\frac{2}{\sqrt{3}}b^{2}\rho ^{\prime }(y)\sech(by)\sin ^{2}(b\tau ). \notag
\end{eqnarray}

We remember that our aim is to get configurations which are periodical in
time. In this sense, it is necessary to impose that the contribution of the
functions $\cos \left( b\tau \right) $ and  $\sin \left( b\tau \right) $ in the right-hand side of the above
partial differential equations should be annulled. Therefore, we can obtain
the functions $\sigma (y)$ and $\rho (y)$, which are given by%
\begin{equation}
\sigma (y)=\sech(by),\text{ }\rho (y)=-\sech(by).  \label{sol1}
\end{equation}

Then, using the solutions (\ref{eq3.6}), (\ref{eq3.7}), and (\ref{sol1}),
the fermionic sector can be writen as%
\begin{equation}
\psi (x,t) \sim \left( 
\begin{tabular}{c}
$\epsilon \sech(b\epsilon x)\cos (b\sqrt{1-\epsilon^{2}}t)$ \\ 
$-\epsilon \sech(b\epsilon x)\sin (b\sqrt{1-\epsilon^{2}}t)$ %
\end{tabular}%
\right) .
\label{oscillino}  \end{equation}


Hence, it was also possible to find oscillon-type configuration in the fermionic sector. This fermionic oscillon is the supersymmetric partner of the oscillon in the bosonic sector.  It is worthy to highlight that this fermionic solution is consistent with the idea of a supersymmetric multiplet. {\bf Indeed, by  applying the transformation (\ref{on_transf_psi}) as a perturbation on the initial configuration (oscillon, Eq. (\ref{e21}), and $\psi = 0$), one generates solutions with non-trivial fermionic sector, which at order$-\epsilon$ read as follows} 
\begin{equation}
\psi (x,t)\simeq \epsilon \, \frac{b^2}{a \sqrt{3}} \, \sech (\epsilon b x) \left( 
\begin{tabular}{c}
$\xi_1 \cos (b\sqrt{1-\epsilon^{2}}t)$ +  $\xi_2 \sin (b\sqrt{1-\epsilon^{2}}t)$ \\ 
$\xi_2 \cos (b\sqrt{1-\epsilon^{2}}t)$ $-$  $\xi_1 \sin (b\sqrt{1-\epsilon^{2}}t)$ %
\end{tabular}%
\right) .
\label{oscillino_via_susy}  \end{equation}

{\bf In the particular case $\xi_2 = 0$, we recover the result in Eq. (\ref{oscillino}) (except for some factors in the amplitude). Here, we re-inforce that Eq. (\ref{oscillino_via_susy}) shall be obtained as a solution to the field equations, if one considers a general solution with both $\sin (b \tau)$- and $\cos (b \tau)$-contributions in Eqs. (\ref{eq3.4}) and (\ref{eq3.5}).

The oscillon and oscillino solutions have the same functional behaviour at order$-\epsilon$, namely,  spatial-localized  and periodical-time dependences, given by  $\sech (\epsilon b x)$ and $\sin (b\sqrt{1-\epsilon^{2}}t)$ or $\cos (b\sqrt{1-\epsilon^{2}}t)$, respectively. However, if we do not restrict ourselves to order$-\epsilon$ in the transformation (\ref{on_transf_psi}), the contributions coming from the non-linear part in $ \xi V_\varphi$ and $i \gamma^\mu \xi \, \partial_\mu \varphi$ will introduce higher powers in $\epsilon$ leading then to different behaviour for the oscillino.   }


\section{Concluding Comments and Further Steps}

\label{sec:csec}

In this work, we have shown that oscillon configurations in $D=1+1$
dimensions, which in this context are also dubbed small amplitude oscillons,
 can be accommodated and duitably in a supersymmetric framework. We have analysed both the bosonic 
and fermionic sectors and noticed that the oscillon and oscillino solutions display similar features at
the $\epsilon-$order; by that, we mean the same oscillation frequency and functional behavior.

 As we have already stated in the Introduction of our work, there are several
reasons that support our proposal of investigating oscillons in connection with
SUSY. The main motivation is based on the fact that, if we wish to consider the
interaction between fermions and oscillons, SUSY provides a natural set-up.
SUSY transformations on oscillon configurations yield the fermionic partners of
the oscillons that lie in the same multiplet. The fermionic excitations induced as SUSY
perturbations on the oscillons propagate and interact with the latter. In turn, the condensates 
of the (fermionic) SUSY partners directly affect the oscillons' propagation, so that, it is an interesting
issue, for a forthcoming contribution, to compute how the oscillon stability changes
by virtue of the presence of the fermionic excitations. 

On the other hand, by taking the viewpoint that fermions are the most elementary matter
excitations in Nature, we intend to go ahead in our endeavour to search for possible
quasi-particle excitations in low-dimensional charged systems that we might identify as
oscillino-like (if the modes are fermionic) or as oscillino condensates, in the case of bosonic
modes. For that, it is mandatory that we focus on the study of charged oscillons in the
framework of an Abelian gauge model to, afterwards, embed the system of charged matter-gauge bosons
in a SUSY context. We shall be reporting on that elsewhere in a forthcoming contribution.

\textbf{Acknowledgments}

RACC would like to thank S\~ao Paulo Research Foundation (FAPESP), grant 2016/03276-5, for financial support. LPRO was supported by the National Council for Scientific and Technological Development (CNPq/MCTIC) through the PCI-DB funds. WP thank CNPq and FAPESP for financial support.


\appendix
\section{Conventions and Superspace Formulation}
\label{Apendix}

Initially, let us fix our conventions. In two-dimensional Minkowski space-time, we adopt the metric $ \eta^{\mu  \nu} = \textrm{diag}(+1,-1) $. One possible choice to satisfy the Clifford algebra, $ \left\{ \gamma^\mu , \gamma^\nu \right\} = 2 \eta^{\mu \nu} $, is given by $ \gamma^0 = \sigma_y $ and $ \gamma^1 = i \sigma_x $, where $\sigma_x $ and $\sigma_y $ denote the usual Pauli matrices. This choice is known as Majorana representation of the gamma matrices.

In order to define a Majorana spinor, we introduce the charge conjugation matrix, $C$, which is antisymmetric $ (C^t = - C )$, unitary $( C^\dagger = C^{-1} )$ and satisfies $ C \gamma^{\mu \, t } C^{-1} = - \gamma^\mu $.  In the Majorana representation, 
we have $ C = - \gamma^0 $. The charge conjugation operation is then defined by $ \psi^c \equiv C \, \bar{\psi}^{\, t} $, where $\bar{\psi} = \psi^\dagger \gamma^0$. A Majorana spinor satisfies the constraint condition $\psi^c = \psi$ and, particularly, in the Majorana representation, we obtain that 
$ \psi^c = \psi \, \Rightarrow \, \psi^\ast = \psi $,  i.e., the spinor has real components.


In what follows, we present the Superspace Formulation. We consider the Supersymmetry $\mathcal{N}=1$ in which the superspace is given by  the coordinates $(x^\mu, \theta)$, where $\theta$ is a Majorana spinor parameter.

We implement the superspace coordinate transformation as a translation, namely,  
\begin{eqnarray}
x'^\mu &=& x^\mu + i \, \bar{\xi} \gamma^\mu \theta \, ,\label{x_transf} \\ 
\theta'_\alpha &=& \theta_\alpha + \xi_\alpha \, , \label{theta_transf}
\end{eqnarray}
where $\alpha=1,2$ and $\xi$ is also a Majorana spinor parameter.

The simplest superfield, including two real scalar fields $(\varphi,F)$ and a Majorana spinor field $(\psi)$, is given by
\begin{equation}
\Phi(x,\theta) = \varphi(x) + \bar{\theta} \, \psi(x) + \bar{\theta} \theta F(x) \, .
\label{superfield} \end{equation}

With the aforementioned transformations, Eqs. \eqref{x_transf} and \eqref{theta_transf}, one may obtain the supersymmetric charge operator $Q$, by considering the variation
\begin{equation}
\delta \Phi \equiv \Phi(x', \theta') - \Phi(x, \theta) = \bar{\xi}  Q \, \Phi(x, \theta) \, ,
\label{Phi_transf} \end{equation} 
which reads
\begin{equation}
Q_{\alpha} = - C_{\alpha \beta} \, \frac{\partial}{\partial \theta_\beta} + i \left( \gamma^\mu \theta \right)_{\alpha} \frac{\partial}{\partial x^\mu}
\label{carga_susy} \, .\end{equation}

%

By comparing Eq. \eqref{Phi_transf} with $\delta \Phi = \delta \varphi  + \bar{\theta} \, \delta \psi + \bar{\theta} \theta \, \delta F$, one may conclude that
\begin{eqnarray}
\delta \varphi &=& \bar{\xi} \psi \, , \label{off_varphi} \\
\delta \psi &=& -i \, \gamma^\mu \xi \, \partial_\mu \varphi + 2 \, F \xi  \, , \label{off_psi} \\ 
\delta F &=& \partial_\mu \left( -\frac{i}{2} \, \bar{\xi} \gamma^\mu \psi \right) \, . \label{off_F}
\end{eqnarray}

Having established the supercharge and transformations, we introduce the supersymmetric covariant derivative,
\begin{equation}
D_{\alpha} = - C_{\alpha \beta} \, \frac{\partial}{\partial \theta_\beta} - i \left( \gamma^\mu \theta \right)_{\alpha} \frac{\partial}{\partial x^\mu} \, ,
\label{cov_der} \end{equation}
which satisfies $ \left\{ D_\alpha , Q_\beta \right\} = 0 $.

We propose the following action in terms of the superfield and covariant derivative,
\begin{equation}
S =  \, \int d^2x \, d^2 \theta \, \left[ - \frac{1}{4} \, \bar{D} \Phi \, D \Phi  + 
\,   V(\Phi) \right] \, .
\end{equation}
where $ \int d^2 \theta \equiv  i \int d \theta_2 d \theta_1 $ and $V(\Phi)$ denotes the superpotential, namely, an arbitrary function of the superfield $\Phi$.

After using some Fierz rearrangements, such as $ \bar{\theta} \psi \, \bar{\theta} \psi = - \frac{1}{2} \, \bar{\theta} \theta \, \bar{\psi} \psi $, and carrying out the Grassmann integral, one can obtain the off-shell Lagrangian density
\begin{equation}
\mathcal{L}_{\textrm{off-shell}} \, = \, \frac{1}{2}(\partial _{\mu
}\varphi )^{2}+\frac{i}{2}\bar{\psi}\gamma ^{\mu }\partial _{\mu }\psi 
+ 2 F^2 + 2 F \, V_\varphi - \frac{\bar{\psi}\psi }{2} V_{\varphi
\varphi } \, .
\label{L_off_shell} \end{equation}

Finally, we notice that $F$ is an auxiliary field. Then, we work out its equation of motion and conclude that 
\begin{equation}
\frac{\delta \mathcal{L}_{\textrm{off-shell}}}{\delta F} = 0 \, \; \Rightarrow \, \; F = - \frac{V_{\varphi}}{2} \, 
\label{eq_F} \, . \end{equation}

Hence, if we substitute this constraint in Eqs. \eqref{L_off_shell} and \eqref{off_varphi}-\eqref{off_psi}, we arrive at the on-shell Lagrangian density \eqref{e1}  and the supersymmetry transformations \eqref{on_transf_varphi}-\eqref{on_transf_psi}, respectively.



\end{document}